\documentclass[preprint,showpacs,showkeys]{revtex4}
\usepackage{mathrsfs}
\usepackage{graphicx}
\usepackage{subfigure}
\usepackage{amssymb}
\usepackage{amsmath}
\usepackage{latexsym}
\usepackage{hyperref}
\usepackage{natbib}

\begin{document}

\date{\today }
\title{Comparison of analytical and numerical methods and the effect of bath
coupling on the quantum decoherence}
\author{Peihao Huang$^{1,2}$}
\email{phhuang@sjtu.edu.cn}
\author{Hang Zheng$^1$}
\author{Keiichiro Nasu$^2$}
\affiliation{$^1$Key Laboratory of Artificial Structures and Quantum Control (Ministry of
Education), Department of Physics, Shanghai Jiao Tong University, Shanghai
200240, China\\
$^2$Solid State Theory Division, Institute of Materials Structure Science, High Energy Accelerator Research Organization (KEK), Tsukuba, Ibaraki 305-0801, Japan}

\begin{abstract}
The dynamics of a qubit in a structured environment is investigated
theoretically. One point of view of the model is the spin-boson model with a
Lorentz shaped spectral density. An alternative view is a qubit coupled to
harmonic oscillator (HO), which in turn coupled to a Ohmic environment. Two
different methods are applied and compared for this problem. One is a
perturbation method based on a unitary transformation. Since the transformed
hamiltonian is of rotating wave approximation (RWA) form, we call it the
transformed rotating wave approximation (TRWA) method. And the other one is
the numerically exact method of the quasi-adiabatic propagator path-integral
(QUAPI) method. TRWA method can be applied from the first point of view. And
the QUAPI method can applied from both points of views. We find that from the
1st point of view QUAPI only works well for large $\Gamma$. Since the memory
time is too long for the practical evaluation of QUAPI when $\Gamma$ is
small. We call this treatment as QUAPI1. And from the 2nd point of view,
QUAPI works well for small $\Gamma$, since the non-adiabatic effect become
more important as $\Gamma$ increases, one need smaller time-step and more steps to obtain
accurate result which also quickly runs out the computational resources.
This treatment is called QUAPI2. We find that the TRWA method works well for
the whole parameter range of $\Gamma$ and show good agreement with QUAPI1
and QUAPI2. On the other hand, we find that the decoherence of the qubit can
be reduced with increasing coupling between HO and bath. This result may be
relevant to the design of quantum computer.
\end{abstract}

\maketitle


\section{Introduction}

Dissipative quantum dynamics is of crucial interest among scientist. Since
the quantum dynamics are always inevitably affected by its environment,
various physical and chemical phenomena are related to the dissipation,
range from the spontaneous emission to electron transfer in molecular, from
qubit decoherence to photon harvest in photosynthesis \cite%
{Nakamura1999,Chiorescu2004,You2005,Engel2007,Lee2007}. Spin-boson model,
the simplest possible model to describe dissipation, offers a comprehensive
way to study the decoherence phenomenon. In the context of the electron
transfer in molecular, spin boson model has been studied intensively over
the past decades, and it shows revival interest among scientists because of
the possible application of quantum computation and information.

For the spin boson model, the environmental property is characterized by the
spectral density $J(\omega )$, which is usually assumed to be a power law
distribution, $J(\omega )\propto \omega ^{s}$, it is called sub-Ohmic when $%
0<s<1$, Ohmic when $s=1$ and super-Ohmic when $s>1$. The most studied case
is the Ohmic spectral density, it describes a case when the dissipation is
the same at all frequencies, which is the case for the many environments. A
physical example corresponds to the Ohmic case is the dissipation in a pure
resistor circuit. For the sub-Ohmic case, it arouse lots of interest
recently, because of the controversy related to the quantum phase transition. \cite%
{Bulla2003,Anders2005,Bulla2005,Vojta2005,Anders2007,Anders2008,Winter2009,Vojta2009,Vojta2010}%
.

In this work, we study the dynamics of a qubit in a structured environment.
Two points views are available for the problem we are interested in. One
point of view is the spin boson model with a Lorentz shaped spectral
density. And the alternative view is a two-level system (TLS) coupled to
harmonic oscillator (HO), which in turn coupled to a Ohmic environment. The
model describes some situations in experiments. For example, a flux qubit is
usually read out by a dc-SQUID with a characteristic plasma frequency, consequently, the environmental noise of the SQUID is also transferred to
the qubit leading to decoherence and dissipation. It
also describes a Cooper-pare box (CPB) coupled to a transmission line resonator or a qubit placed in a leaky cavity.

From the first point of view, many tradition treatment of spin-boson model
can be applied, some of them are largely numerical, such as the quantum
Monte Carlo \cite{Egger1992,Egger1994,Egger2000}, real-time renormalization
group \cite{Keil2001}, quasi-adiabatic path integrals (QUAPI) \cite%
{Makarov1994,Makri1995a,Makri1995b,Thorwart2004}, flow equation
renormalization (FER) \cite%
{costi1996,kehrein1997,Grifoni1999,Kleff2003,Wilhelm2004,Kleff2004} and
numerical renormalization group (NRG) \cite%
{Anders2005,Anders2006,Anders2007,Anders2008,Bulla2008}, others are mainly
analytical such as the non-interacting blip approximation (NIBA) \cite%
{Leggett1987,Weiss1999,Wilhelm2004,Nesi2007a,Nesi2007b}, rigorous Born approximation \cite%
{Grifoni1999,DiVincenzo2005} or Bloch-Redfield \cite%
{Redfield1965,Hartmann2000}, and the transformed rotating-wave approximation
(TRWA) method \cite{Zheng2004,Wang2010,Huang2008,Gan2010a}. Till now, for
this particular spectral density, it has been studied by FER \cite%
{Kleff2003,Wilhelm2004,Kleff2004}, NIBA \cite{Wilhelm2004,Nesi2007a},
Bloch-Redfield \cite{Wilhelm2004} and TRWA \cite{Huang2008,Gan2010a}. And from
the second point of view, it has been studied by QUAPI \cite{Thorwart2004},
the van Vleck perturbation theory together with a Born-Markov master
equation (VVBM) \cite{Hausinger2008}, Redfield equation based on a single
mode TRWA \cite{Brito2008}. Although it has been studied intensively, the
most studied case is the weak HO-bath coupling. While in this work, we
explore the qubit decoherence in both the weak and strong HO-bath coupling
regimes.

One commonly believed concept is that temperature and the coupling to noisy
bath only play a negative role in preserving the qubit coherence. However,
it is pointed out in Ref. \cite{Montina2008} that the temperature can help
the coherence when the qubit is coupled to a TLF (or spin-boson)
environment. We would like to ask: Is it possible to reduce the decoherence
by increasing the system-bath coupling? In this paper, we examine the
coupling effect on the quantum decoherence, we study aforementioned problem
from the 1st point of view by the TRWA method in both the weak and strong
HO-bath coupling regimes. To check the result, we also use the numerically
exact method of the quasi-adiabatic propagator path-integral (QUAPI) method.
We find that from the 1st point of view QUAPI only works well for large $%
\Gamma $. Since the memory time is too long for the practical evaluation of
QUAPI when $\Gamma $ is small. We call this treatment as QUAPI1. And from
the 2nd point of view, QUAPI works well for small $\Gamma $, since the
non-adiabatic effect of the Ohmic bath become more important as $\Gamma $
increases, consequently, smaller time-step is required, thus, for the same
memory time, one need more steps to obtain accurate result which also
quickly runs out the computational resources. This treatment is called
QUAPI2. We find that the TRWA method works well for the whole parameter
range of $\Gamma $ and show good agreement with QUAPI1 and QUAPI2. On the
other hand, we find that the decoherence of the qubit can be reduced with
increasing HO-bath coupling when the HO-bath coupling is larger than the
qubit-HO coupling.

The paper is organized as follows. In sec. \ref{secii}, we present
explicitly the two points of views of the problem we are interested in. In
sec. \ref{seciii} and \ref{seciv}, we briefly introduce the TRWA and the
QUAPI method, and discuss how to adopt these methods. In sec. \ref{secv},
the coupling effect on the quantum coherence is discussed and a comparison
of the TRWA with QUAPI is given. Finally, a brief conclusion is presented in
Sec. \ref{secvi}.

\section{Two different points of views of the problem}

\label{secii}

The 1st point of view of the problem is the spin-boson model (un-biased
case)
\begin{equation}
H=-\frac{\Delta}{2}\sigma_x+\sum_k{\omega_k}b^{\dagger}_kb_k+\frac{1}{2}
\sigma_z\sum_kg_k(b_k^{\dagger}+b_k),  \label{Hamiltonian3}
\end{equation}
where $\sigma_x$ and $\sigma_z$ are the pauli matrices, $\Delta$ is the gap
of of the qubit, ${b}_k$ (or ${b}_k^\dag$) are the annihilation (or
creation) operators of the bath. And the spectral density is Lorentzian,
\begin{equation}  \label{J}
J(\omega)\equiv\sum_kg_k^2{\delta} (\omega-\omega_k)=\frac{%
2\alpha\omega\Omega^4}{(\Omega^2-\omega^2)^2+(2\pi\Gamma\omega\Omega)^2}.
\end{equation}
This spectral density possesses a shape peak at position $\Omega$ especially
when $\Gamma$ is small, which may challenge the conventional method of the
spin boson model. In the limit of $\Omega\to\infty$, $J(\omega)$ is reduced
to the Ohmic spectral density.

The above model can be exactly mapped to a model where the qubit is
dissipated by a multi-mode Ohmic bath via a harmonic oscillator, which is
the 2nd point of view of this problem \cite{Garg1985,Tian2002}. The
Hamiltonian reads ($\hbar =1$)
\begin{eqnarray}
H &=&-\frac{\Delta }{2}\sigma _{x}+{\Omega }B^{\dagger }B+\sum_{k}\widetilde{%
\omega }_{k}\tilde{b}_{k}^{\dagger }\tilde{b}_{k}  \notag \\
&+&(B^{\dagger }+B)\bigg[g_{0}\sigma _{z}+\sum_{k}\kappa _{k}(\widetilde{b}%
_{k}^{\dagger }+\widetilde{b}_{k})\bigg]+(B^{\dagger }+B)^{2}\sum_{k}\frac{%
\kappa _{k}^{2}}{\widetilde{\omega }_{k}},  \label{Hamiltonian2}
\end{eqnarray}%
where, $B$ (or $B^{\dag }$) are the annihilation (or creation) operators of
the HO, $\widetilde{b}_{k}$ (or $\widetilde{b}_{k}^{\dagger }$) are the
annihilation (or creation) operators of the corresponding bath mode. The
last term is the counter-term, which cancels the additional contribution due
to the coupling of the HO to the bath \cite{Weiss1999,Costa2000}. The
corresponding spectral density is of Ohmic form
\begin{equation}
\tilde{J}(\omega )\equiv \sum_{k}\kappa _{k}^{2}{\delta }(\omega -\widetilde{%
\omega }_{k})=\Gamma \omega e^{-\omega /\omega _{c}}
\end{equation}%
where $\omega _{c}$ is the cut-off frequency (throughout this work, we use
the value $\omega _{c}=20\Omega $). The relation between $g_{0}$ and $\alpha
$ follows as $g_{0}={\Omega }\sqrt{\frac{\alpha }{8\Gamma }}$. From this 2nd
point of view, the over all system TL plus HO can be considered as the
system which is then dissipated by an Ohmic bath.

\section{Transformed Rotating-wave approximation method}

\label{seciii}

The transformed rotating-wave approximation method is developed by one of
the author and then applied to a sequence of problems \cite%
{Zheng2004,Wu2005,Zhu2005,Cao2007a,Cao2007b,Lu2007,Cao2008,Huang2008,Zheng2008,Cao2009,Gan2009,Lu2009,Wang2009,Gan2010a,Gan2010b,Wang2010,Yang2010}
Our starting point is the SBM with the structured spectral density. A
unitary transformation is first applied to the SBM Hamiltonian, $H^{\prime
}=\exp (S)H\exp (-S)$, with the generator $S\equiv \sum_{k}\frac{g_{k}}{%
2\omega _{k}}\xi _{k}(b_{k}^{\dag }-b_{k})\sigma _{z}$. \bigskip The
transformed Hamiltonian can be decomposed into three parts:
\begin{eqnarray}
H_{0}^{\prime } &=&-\frac{\sigma _{x}}{2}\eta \Delta +\sum\limits_{k}{\omega
_{k}}b_{k}^{\dagger }b_{k}-\sum\limits_{k}\frac{g_{k}^{2}}{4\omega _{k}}\xi
_{k}(2-\xi _{k}), \\
H_{1}^{\prime } &=&\frac{\sigma _{z}}{2}\sum\limits_{k}g_{k}(1-\xi
_{k})(b_{k}^{\dagger }+b_{k})-\frac{i\sigma _{y}}{2}\eta \Delta {X},
\label{formerH_1'} \\
H_{2}^{\prime } &=&-\frac{\sigma _{x}}{2}\Delta \left\{ \cosh {X}-\eta
\right\} -\frac{i\sigma _{y}}{2}\Delta \bigg\{\sinh {X}-\eta {X}\bigg\},
\label{H_2'}
\end{eqnarray}%
where, $X\equiv {\sum_{k}\frac{g_{k}}{\omega _{k}}\xi _{k}(b_{k}^{\dag
}-b_{k})}$ and $\eta $ is the thermodynamic average of $\cosh {X}$. In the
limit of zero temperature it is,
\begin{equation}
\eta =\exp \left[ -\sum\limits_{k}\frac{g_{k}^{2}}{2\omega _{k}^{2}}\xi
_{k}^{2}\right] .  \label{eta}
\end{equation}%
Obviously, $H_{0}^{\prime }$ can be solved exactly since the spin and bosons
are decoupled in $H_{0}^{\prime }$. $\eta \Delta $ gives a rough
approximation of the renormalized qubit frequency and $(\eta -1)\Delta $ is
the corresponding Lamb shift of the qubit due to the coupling to the bath.
The eigenstate of $H_{0}^{\prime }$ can be expressed as direct product: $%
|s\rangle |\{n_{k}\}\rangle $, where $|s\rangle $ is the eigenstate of $%
\sigma _{x}$ and $|\{n_{k}\}\rangle $ is the eigenstate of phonons, which
means that there are $n_{k}$ phonons for mode $k$. Therefore, the ground
state of $H_{0}^{\prime }$ is given by : $|g_{0}\rangle =|\,s_{1}{\rangle }%
|\{0_{k}\}\rangle ,$ where $|\,s_{1}{\rangle }$ is the lower eigenstate of
spin and $|\{0_{k}\}\rangle $ stands for the vacuum state of the bosons.

The choice of $\eta $ in Eq.~(\ref{eta}) insures $H_{2}^{\prime }$ contains
only the terms of two-boson and multi-boson non-diagonal transitions and its
contribution to physical quantities is $(g_{k}^{2})^{2}$ and higher.
Therefore, $H_{2}^{\prime }$ can be omitted in the following discussion. If
we let $H_{1}^{\prime }|g_{0}\rangle =0$, which ensures $H_{1}^{\prime }$ to
be small and more suitable for the subsequent perturbation treatment, then
the parameters $\xi _{k}$'s are determined as,
\begin{equation}
\xi _{k}=\frac{\omega _{k}}{\omega _{k}+\eta \Delta }.  \label{xi_k}
\end{equation}%
Consequently, $H_{1}^{\prime }$ is transformed to the rotating-wave form,
\begin{equation*}
H_{1}^{\prime }=\sum_{k}V_{k}(b_{k}^{\dagger }\sigma _{-}+b_{k}\sigma _{+}),
\end{equation*}%
where $V_{k}=\eta {\Delta }g_{k}\xi _{k}/\omega _{k}={g_{k}\eta \Delta }/{%
(\omega _{k}+\eta \Delta )}$ and $\sigma _{\pm }\equiv (\sigma _{z}{\mp }%
i\sigma _{y})/2$. Actually, some effect of the anti-rotating-wave terms has
been taken into account in $H_{1}^{\prime }$ by the unitary transformation
which is embodied in the renormalized coupling constant $V_{k}$. Note that $%
\xi _{k}\sim 1$ if the boson frequency $\omega _{k}$ is larger than the
renormalized tunneling $\eta \Delta $, but $\xi _{k}\ll 1$ for $\omega
_{k}\ll \eta \Delta $. Since the transformation generated by $S$ is a
displacement of bosons, physically, one can see that high-frequency bosons ($%
\omega _{k}>\eta \Delta $) follow the tunneling particle adiabatically
because the displacement is $g_{k}\xi _{k}/\omega _{k}\sim {g_{k}/\omega _{k}%
}$. However, bosons of low-frequency modes $\omega _{k}<\eta \Delta $ in
general are not always in equilibrium with the tunneling particle, hence the
particle moves in a retarded potential arising from the low-frequency modes.
When the non-adiabatic effect dominates, $\omega _{k}\ll \eta \Delta $, the
displacement $\xi _{k}\ll 1$.

The total density matrix (system+environment) $\chi ^{\prime }(t)$ obeys the
Liouville-von-Neumann equation,
\begin{equation}
\frac{d}{dt}\tilde{\chi ^{\prime }}(t)=-i[\tilde{H_{1}^{\prime }}(t),\tilde{%
\chi ^{\prime }}(t)],
\end{equation}%
where the tildes denote operators in the interaction picture with respect to
$H_{0}^{\prime }$. Iterating up to the second order and tracing out the
environmental degrees, one get the master equation within the Born
approximation,
\begin{equation}
\frac{d}{dt}\tilde{\rho ^{\prime }}(t)=-i\mathrm{Tr}_{B}[\tilde{%
H_{1}^{\prime }}(t),\rho _{B}^{\prime }\otimes \tilde{\rho ^{\prime }}%
(0)]-\int_{0}^{t}dt^{\prime }\mathrm{Tr}_{B}[\tilde{H_{1}^{\prime }}(t),[%
\tilde{H_{1}^{\prime }}(t^{\prime }),\rho _{B}^{\prime }\otimes \tilde{\rho
^{\prime }}(t^{\prime })]],  \label{MasterEquation1}
\end{equation}%
where $\tilde{\rho ^{\prime }}$ is the reduced density matrix $\tilde{\rho
^{\prime }}(t)=$Tr$_{B}\left[ \tilde{\chi ^{\prime }}(t)\right] $, and $%
\tilde{\chi ^{\prime }}(t)$ is replaced by an approximate factorized density
matrix $\tilde{\chi ^{\prime }}(t)\approx \rho _{B}^{\prime }\otimes \tilde{%
\rho ^{\prime }}(t)$. The environment is usually assumed to remain in
thermal equilibrium $\rho _{B}^{\prime }={e^{-\beta H_{B}^{\prime }}}/{%
\mathrm{Tr}e^{-\beta H_{B}^{\prime }}}$ with $H_{B}^{\prime }=\sum\limits_{k}%
{\omega _{k}}b_{k}^{\dagger }b_{k}$, which is justified when the environment
is 'very large' and the coupling 'weak' ($V_{k}\ll \Delta ,\Omega )$, so
that the back-action of the system onto the environment can be neglected.
Substitute $H_{1}^{\prime }$ into the master equation, we get
\begin{equation}
\frac{\partial \tilde{\rho ^{\prime }}(t)}{{\partial }t}=-iTr_{B}[H_{1}^{%
\prime },\tilde{\rho ^{\prime }}(t)]-\sum_{k}V_{k}^{2}\int_{0}^{t}d\,t^{%
\prime }\tilde{X}(t,t^{\prime })  \label{ME_I}
\end{equation}%
where,
\begin{eqnarray*}
\tilde{X}(t,t^{\prime }) &\equiv &n_{k}[\sigma _{-}(t)\sigma _{+}(t^{\prime
})\tilde{\rho ^{\prime }}(t^{\prime })-\sigma _{+}(t^{\prime })\tilde{\rho
^{\prime }}(t^{\prime })\sigma _{-}(t)]e^{i\omega _{k}(t-t^{\prime })} \\
&+&n_{k}[\tilde{\rho ^{\prime }}(t^{\prime })\sigma _{-}(t^{\prime })\sigma
_{+}(t)-\sigma _{+}(t)\tilde{\rho ^{\prime }}(t^{\prime })\sigma
_{-}(t^{\prime })]e^{-i\omega _{k}(t-t^{\prime })} \\
&+&(n_{k}+1)[\sigma _{+}(t)\sigma _{-}(t^{\prime })\tilde{\rho ^{\prime }}%
(t^{\prime })-\sigma _{-}(t^{\prime })\tilde{\rho ^{\prime }}(t^{\prime
})\sigma _{+}(t)]e^{-i\omega _{k}(t-t^{\prime })} \\
&+&(n_{k}+1)[\tilde{\rho ^{\prime }}(t^{\prime })\sigma _{+}(t^{\prime
})\sigma _{-}(t)-\sigma _{-}(t)\tilde{\rho ^{\prime }}(t^{\prime })\sigma
_{+}(t^{\prime })]e^{i\omega _{k}(t-t^{\prime })}
\end{eqnarray*}

The master equation Eq.~(\ref{ME_I}) which is a $2\times 2$ matrix equation
can be solved exactly by the Laplace transform since the convolution theorem
can be applied to the equation of each matrix element. Here, for simplicity,
we only present the comparatively brief expression. Suppose the system is in
the upper eigenstate of $\sigma _{z}$ at time $t=0$. At the zero
temperature, the population difference $P(t)\equiv \langle \sigma
_{z}(t)\rangle \equiv \mathrm{Tr}_{A}(\sigma _{z}\rho (t))$ is evaluated as
\begin{equation}
P(t)=\frac{1}{\pi }\int_{0}^{\infty }d\omega \,\frac{\gamma (\omega )\cos
(\omega t)}{\left[ \omega -{\eta }\Delta -R(\omega )\right] ^{2}+\gamma
^{2}(\omega )}.  \label{P(t)_int}
\end{equation}%
$R(\omega )$ and $\gamma (\omega )$ in Eq.~(\ref{P(t)_int}) are the real and
imaginary parts of $\sum_{k}V_{k}^{2}/(\omega -i0^{+}-\omega _{k})$
\begin{eqnarray}
R(\omega ) &=&\int_{0}^{\infty }\,d\omega ^{\prime }\frac{(\eta \Delta
)^{2}J(\omega ^{\prime })}{(\omega ^{\prime }+\eta \Delta )^{2}(\omega
-\omega ^{\prime })},  \label{R} \\
\gamma (\omega ) &=&{\pi }J(\omega )(\eta \Delta )^{2}/(\omega +\eta \Delta
)^{2}.  \label{r}
\end{eqnarray}

\section{Quasi-adiabatic path-integral method}

\label{seciv}

The QUAPI method is a numerical scheme based on a exact methodology \cite%
{Makarov1994,Makri1995a,Makri1995b,Thorwart2004}. The starting point of
QUAPI method is the generic system-bath Hamiltonian
\begin{equation*}
H=H_{0}+\sum_{j}\frac{P_{j}^{2}}{2m_{j}}+\frac{1}{2}m_{j}\omega
_{j}^{2}(Q_{j}-c_{j}s/m_{j}\omega _{j}^{2})^{2}.
\end{equation*}%
where, $H_{0}$ is the Hamiltonian for the bare system, $s$ is the system
coordinate, and $Q_{j}$ are harmonic bath coordinates which are linearly
coupled to the system coordinate. The characteristics of the bath are
captured in the spectral density function
\begin{equation}
J(\omega )=\frac{\pi }{2}\sum_{j}\frac{c_{j}^{2}}{m_{j}\omega _{j}}\delta
(\omega -\omega _{j}).
\end{equation}%
The reduced density matrix of the system evolve as $\rho (s^{\prime \prime
},s^{\prime },t)=\mathtt{Tr}_{bath}\left\langle s^{\prime \prime
}\right\vert e^{-iH_{0}t}\rho (0)e^{iH_{0}t}\left\vert s^{\prime
}\right\rangle $. If the path integral representation is discretized by $N$
time steps of length $\Delta t=t/N$ and the initial density matrix is
assumed to be $\rho (0)=\rho _{s}(0)\rho _{bath}(0)$, the reduced density
matrix takes the form%
\begin{eqnarray}
&&\rho (s^{\prime \prime },s^{\prime
},t)=\sum_{s_{N-1}^{+}}\sum_{s_{N-1}^{-}}\cdots
\sum_{s_{1}^{+}}\sum_{s_{1}^{-}}\sum_{s_{0}^{+}}\sum_{s_{0}^{-}}\left\langle
s^{\prime \prime }\right\vert e^{-iH_{0}\Delta t}\left\vert
s_{N-1}^{+}\right\rangle \cdots \left\langle s_{1}^{+}\right\vert
e^{-iH_{0}\Delta t}\left\vert s_{0}^{+}\right\rangle  \notag \\
&&\left\langle s_{0}^{+}\right\vert \rho _{s}(0)\left\vert
s_{0}^{-}\right\rangle \left\langle s_{0}^{-}\right\vert e^{iH_{0}\Delta
t}\left\vert s_{1}^{-}\right\rangle \cdots \left\langle
s_{N-1}^{-}\right\vert e^{iH_{0}\Delta t}\left\vert s^{\prime }\right\rangle
I(s_{0}^{+},s_{0}^{-},s_{1}^{+},s_{1}^{-},\cdots
,s_{N-1}^{+},s_{N-1}^{-},s^{\prime \prime },s^{\prime },\Delta t)  \notag \\
&&  \label{rhoss}
\end{eqnarray}%
where the discrete variable representation (DVR) is used, the symbol $%
s_{k}^{\pm }$ ($k=0.....N-1$) denotes the system coordinate at the time $%
k\Delta t$ on the forward and backward discretized Feynman path. $\left\vert
s_{k}^{\pm }\right\rangle $ ($k=0.....N-1$) are the eigenstates of the
system coordinate operator $s$. If a symmetric splitting of the
time-evolution operator is employed $e^{-iH\Delta t}=e^{-iH_{env}\Delta
t/2}e^{-iH_{0}\Delta t}e^{iH_{env}\Delta t/2}$ with $H_{env}=H-H_{0}$, the
corresponding influence functional reads%
\begin{eqnarray}
&&I(s_{0}^{+},s_{0}^{-},s_{1}^{+},s_{1}^{-},\cdots
,s_{N-1}^{+},s_{N-1}^{-},s^{\prime \prime },s^{\prime },\Delta t)  \notag \\
&=&\mathtt{Tr}_{bath}\left[ e^{-iH_{env}(s^{\prime \prime })\Delta
t/2}e^{-iH_{env}(s_{N-1}^{+})\Delta t}\cdots e^{-iH_{env}(s_{0}^{+})\Delta
t/2}\right.  \notag \\
&&\left. \times \rho _{bath}(0)e^{iH_{env}(s_{0}^{-})\Delta t/2}\cdots
e^{iH_{env}(s_{N-1}^{-})\Delta t}e^{iH_{env}(s^{\prime })\Delta t/2}\right] ,
\end{eqnarray}%
One can find that the equilibrium position of the bath mode is adiabatically
displaced along the system coordinate. If $H_{0}$ provides a reasonable
zeroth-order approximation to the dynamics, the quasi-adiabatic propagator
is accurate for fairly large time steps. That is the quasi-adiabatic
partitioning is a good representation when the bath property is mainly
adiabatic, where the bath can keep up with the motion the system quickly.
And the discrete path is to take into account of the non-adiabatic effect.
For most of case, the quasi-adiabatic partitioning is reasonable especially
when the system bath coupling is not strong. Therefore, the QUAPI
discretization permits fairly large time steps when the adiabatic bath
dominates the system dynamics. If the bath is purely adiabatic, even no
discretization is needed. In the continuous limit (that is for $\Delta {t}%
\rightarrow 0,N\rightarrow \infty $) the influence functional has been
calculated by Feynman and Vernon
\begin{eqnarray}
I &=&exp\left\{ -\frac{1}{\hbar }\int_{0}^{t}dt^{\prime }\int_{0}^{t^{\prime
}}dt^{\prime \prime }\left[ s^{+}(t^{\prime })-s^{-}(t^{\prime })\right]
\right.  \notag \\
&&\times \left[ \alpha (t^{\prime }-t^{\prime \prime })s^{+}(t^{\prime
\prime })-\alpha ^{\ast }(t^{\prime }-t^{\prime \prime })s^{-}(t^{\prime
\prime })\right]  \notag \\
&&\left. -\frac{i}{\hbar }\int_{0}^{t}dt^{\prime }\sum_{j}\frac{c_{j}^{2}}{%
2m_{j}\omega _{j}^{2}}\left[ s^{+}(t^{\prime })^{2}-s^{-}(t^{\prime })^{2}%
\right] \right\}  \label{I_FeynmanVernon}
\end{eqnarray}%
where $\alpha (t)$ is the bath response function, which can be expressed in
terms of the spectral density as
\begin{equation}
\alpha (t)=\frac{1}{\pi }\int_{0}^{\infty }{d\omega }J(\omega )\left[ \coth
\left( \frac{\beta \omega _{j}\hbar }{2}\cos (\omega _{j}t)-i\sin (\omega
_{j}t)\right) \right] .  \label{alphat}
\end{equation}%
The last term in Eq.(\ref{I_FeynmanVernon}) arises from the "counter-terms"
which are grouped with the bath Hamiltonian in the quasi-adiabatic splitting
of the propagator. With the quasi-adiabatic discretization of the path
integral, the influence functional, Eq. \ref{I_FeynmanVernon}, takes the form%
\begin{equation}
I=exp\left\{ -\frac{1}{\hbar }\sum\limits_{k=0}^{N}\sum\limits_{k^{\prime
}=0}^{k}\left[ s_{k}^{+}-s_{k}^{-}\right] \left[ \eta _{kk^{\prime
}}s_{k^{\prime }}^{+}-\eta _{kk^{\prime }}^{\ast }s_{k^{\prime }}^{-}\right]
\right\} ,  \notag
\end{equation}%
where $s_{N}^{+}=s^{\prime \prime }$ and $s_{N}^{-}=s^{\prime }$. The
coefficients $\eta _{kk^{\prime }}$ can be obtained by substituting the
discretized path into the Feynman-Vernon expression Eq.(\ref{I_FeynmanVernon}%
), which is given in Ref. \cite{Makri1995a}.

The QUAPI method is essentially a tensor multiplication scheme, which
exploits the observation that for environments characterized by broad
spectra the response function $\alpha (t)$ decays within a finite time
interval. From the expression of the Feynman and Vernon influence funcitonal
Eq. (\ref{rhoss}), one can see that $\alpha (t)$ characterizes nonlocal
interactions, which connects system coordinate $s(t^{\prime })$ with $%
s(t^{\prime \prime })$. The path $s^{\pm }(t^{\prime })$ at time $t^{\prime
} $ is connected to the all the paths $s^{-}(t^{\prime \prime })$ at earlier
times, which makes the evaluation of Eq. (\ref{rhoss}) a hard task. However,
for a bath with a broad spectral density, such as a power law distribution
of the spectral density, $\alpha (t)$ has the finite memory, the memory
length typically extending over only a few time slices when the
quasi-adiabatic propagator is used to discretize the path integral. After
discarding the negligible "long-distance interaction" with $t^{\prime
}-t^{\prime \prime }>\Delta k_{\max }\Delta t$ (or $k-k^{\prime }>\Delta
k_{\max }$), the resulting path integral can be evaluated iteratively by
multiplication of a tensor of rank $2\Delta k_{max}$. In other words, there
exists an augmented reduced density tensor of rank $\Delta k_{max}$ that
obeys Markovian dynamics. The details of the multiplication scheme is
discussed to a great extent in the literature, here we only present the
essential parameters and mention briefly how to adopt it to our specific
problem \cite{Makarov1994,Makri1995a,Makri1995b,Thorwart2004}.

\subsection{QUAPI1: treatment from the 1st point of view}

The Quapi method is based on an exact methodology, and for most conventional
bath, such as bath with power law distributed spectral density, QUAPI work
efficiently and converge easily. However, for the spectral density studied
here, it pose challenge to the application of the QUAPI method, since the
memory length is too long for the practical implementation of the QUAPI when
$\Gamma $ is small. The response function is depicted in Fig.~1, one can
find that because of the characteristic peak frequency $\Omega $ of the
spectral density, the response function possess a coherent oscillation with
frequency $\Omega $, and possesses very long memory time. However, when $%
\Gamma $ becomes large, the damping of the kernel becomes significant, and
the response function quickly damp to the zero. which enables the
application of the QUAPI. Here we discuss briefly the parameters used in the
QUAPI1 method:

(i)The first parameter time-step $\Delta {t}$ used for the quasi-adiabatic
splitting of the path-integral. The memory time of the non-Markovian steps
used by QUAPI is $\Delta {k_{max}}\Delta {t}$. The stability of the
iterative density matrix propagation ensures the choices of $\Delta {t}$, it
should not be too big nor too small, since the non-adiabatic effect requires
more splitting of the path integral, that is smaller $\Delta {t}$. Whereas,
since the memory length $\Delta {k_{max}}\Delta {t}$ is usually a fixed
value for a particular bath, QUAPI method prefers larger $\Delta {t}$, and
consequently smaller $\Delta {k_{max}}$ in consideration of the
numerical efficiency (note that the algorithm scales exponentially with $\Delta
k_{\max }$, also see the discussion of the second parameter $\Delta k_{\max }$).
Therefore, we should choose appropriate $\Delta {t}$ to take into account
both the non-adiabatic effect which prefer smaller time splitting and the
non-Markov effect which prefer long memory time, typically, we choose $%
\Delta t$ around $\frac{2\pi }{20\Delta }$, that is to choose tens of
fraction of the cycle time of the bare system dynamics.

(ii) The second parameter is the memory steps $\Delta {k_{max}}$. If $\Delta {k_{max}}\leq 1$, the dynamics is purely Markovian. If the non-locality extends over longer time,
terms with $\Delta {k_{max}}>1$ have to be included to obtain accurate
results. In order to acquire converge result, in the practical
implementation of QUAPI, one usually need to choose $\Delta k_{\max }$ large
enough so that the response function reduces to negligible value within the
length of $\Delta {k_{max}}\Delta {t}$. However this is a hard task, Since
augmented propagator tensor $A^{(\Delta {k_{max}})}$ is a vector of
dimension $(M^{2})^{\Delta {k_{max}}}$ ($M$ is the system dimension which is
two here), and the corresponding tensor propagator $T^{(2\Delta {k_{max}})}$
is a matrix of dimension $(M^{2})^{2\Delta {k_{max}}}$, the QUAPI scheme
scales exponentially with the parameter $\Delta k_{\max }$. Thus one can not
proceed the QUAPI calculation with very large $\Delta k_{\max }$, and
usually $\Delta k_{\max }$ is chosen less than 10 for $M=2$, and even
smaller for larger $M$.

In summary, one have to select appropriate $\Delta {t}$ and $\Delta {k_{max}}
$ to achieve stable and accurate result. As discussed in \cite%
{Makri1995a,Makri1995b} for the SBM problem with the ohmic bath, with the
choice of $\Delta {t=}0.25/\Omega $ ($\Omega $ is the tunneling rate of the
bare system there), only $\Delta {k_{max}}=5,7$ reaches stability for zero
bias (as shown in Fig.~2 in \cite{Makri1995a}) and only $\Delta {k_{max}}=7,9
$ reaches the long time limit for nonzero bias (as shown in Fig.~5 in \cite%
{Makri1995a}). For our system, it is possible for the implementation of
QUAPI only when $\Gamma $ is large enough, since the memory time is too long
to implement QUAPI when $\Gamma $ is small.

\subsection{QUAPI2: treatment from the 2nd point of view}

As discussed above, the direct application of QUAPI method is impossible
when $\Gamma $ is small, how ever, we can tackle this problem from the
second point of view, that is qubit is first coupled to a harmonic
oscillator, which itself coupled to a Ohmic bath. Since the memory length of
the Ohmic bath extend over only a few slices of the time steps. It enable the use
of the QUAPI if we take the over all system of qubit-HO as the $H_{0}$ and
the Ohmic bath as the environment. It work well especially for the weak $%
\Gamma $ in this case, since the non-adiabatic effect is weak for the small
coupling and $H_{0}$ provides a reasonable zeroth-order approximation to the
dynamics. Therefore $\Delta {t}$ can be fairly large, and the result
converges quickly as $\Delta {k_{max}}$ increases.

However, since the qubit-HO system possess an infinite number of energy
levels, it prohibit the direct use of the QUAPI. We have to truncate the
qubit-HO system into a smaller sub-space for the practical implementation of
the QUAPI method. Similarly to the Ref. \cite{Thorwart2004}, we first
diagonalize the qubit-HO system in a large $N$ dimensional space and then
only preserve the lowest $M$ energy eigenstates, consequently the system
coordinate operator $X$ is also truncated to the $M$ dimension operator $%
X_{M}$ in the eigen energy representation. Then we diagonalize the $X_{M}$
to transform into the DVR basis. Since now the system is only $M$
dimensional, the implementation of QUAPI method becomes feasible.

Parameters in QUAPI2: (i) $\Delta {t}$ and (ii) $\Delta {k_{max}}$ are the
same as QUAPI1. And two additional parameter appears in QUAPI2:

(iii) One parameter is the dimension $N$ of the Hilbert space of the
qubit-HO system. The Hamiltonian of the qubit-HO system can be numerically
diagonalized in the $N$ dimension of the Hilbert space. $N$ is kept fixed as $N=400$ in
our calculation, since it is big enough dimension for the
diagonalization of the qubit-HO system.

(iv) Meanwhile, we have employed a second parameter $M$, which is the lowest
energy sub-space of the dimension $N$ of the Hilbert space of the qubit-HO
system. We first diagonalize the qubit-HO space in the larger dimension $N$
of the Hilbert space, to get more accurate low energy eigenstates and
calculate the physical quantities in the $M$ dimension subspace with less
numerical effort. Here we should choose larger $M$ for stronger
qubit-HO coupling $g_{0}$.

\section{Results and discussion}

\label{secv}

According to Eq.~(\ref{J})-(\ref{eta}) and Eq.~(\ref{P(t)_int})-(\ref{r}), $%
P(t)$ is obtained according to TRWA method. Here we report $P(t)$ as a
function of time in Fig.~2 and Fig.~3 for the off-resonance case ($\Delta =0.1\Omega$), and in Fig.~4 and Fig.~5 for the on-resonance case ($\Delta =\Omega$). For the off-resonance case as shown in Fig.~2 and Fig.~3, the decoherence is always enhanced by increasing the HO-boson coupling $\Gamma$ no matter that $\pi \Gamma \Delta $ is larger or smaller than $g_{0}$. However, for the on-resonance case, the decoherence is enhanced with increasing $\Gamma $ when $\pi \Gamma \Delta>g_{0}$ as shown in Fig.~4, Whereas reduced with $\Gamma$ when $\pi \Gamma \Delta<g_{0}$ as shown in Fig.~5.

To check the peculiar results, we also calculate the population difference $P(t)$ by QUAPI
method. For $\pi \Gamma \Delta >g_{0}$ we do the QUAPI from the first point
of view (QUAPI1) and the result is reported in Fig.~3 and Fig.~5, the time splitting is
set fixed as $\Delta t=0.6/\Delta$ with varying memory steps $\Delta
k_{\max }$. In Fig.~5, one can find that when $\Gamma $ is large, i.e. $\Gamma =0.4,0.5$%
, it is easy for the calculation to converge and the result is good enough within $%
\Delta k_{\max }=3$. As $\Gamma $ decreases, it become harder for the
evaluation because of the long memory time. In order to converge to our analytical
result, it needs $\Delta k_{\max }=5$ for $\Gamma =0.3$, $\Delta k_{\max }=7
$ for $\Gamma =0.2$, and even higher $\Delta k_{\max }$ for $\Gamma =0.1$
(which almost runs out our numerical resources). For even smaller $\Gamma $%
, such as $\Gamma =0.02$, the evaluation from this point of view is
practically impossible. Therefore, for $\pi \Gamma \Delta <g_{0}$ we do the
QUAPI from the second point of view (QUAPI2), which is reported in Fig.~2 and Fig.~4.
Here the time splitting is set fixed as $\Delta t=0.15/\Delta$ in Fig.~2 and $\Delta t=0.3/\Delta$ in Fig.~4, the
dimension for diagonalization of TL-HO system is $N=400$ and truncated
dimension for QUAPI2 is $M=2$ in Fig.~2 and $M=6$ in Fig.~4. One can see that all these results converges to our
TRWA results.

To understand the qubit behavior, we can explore the damping rate
according to the TRWA method. One can find that in the near resonance case ($%
\Delta \approx \Omega $), level repulsion occurs, two characteristic
frequencies dominate the qubit dynamics. When the HO-bath coupling is weak,
the result should agree with that of the Jaynes-Cummings model, that is the
peak frequency $\omega _{p}\approx \Delta \pm g_{0}$ \cite{Jaynes1963}.
Therefore, from Eq. (\ref{r}), we get
\begin{equation}
\gamma _{p}\propto \frac{g_{0}^{2}\Gamma }{g_{0}^{2}+(\pi \Gamma \Delta )^{2}%
},  \label{gammap}
\end{equation}%
where we have approximated $\eta \approx 1$, $\omega _{p}+\eta \Delta
\approx 2\Delta $ and $\omega _{p}+\Omega \approx 2\Delta $. Therefore, one
expect $\gamma _{p}\propto \Gamma $ for $\pi \Gamma \Delta {\ll }g_{0}$ and $%
\gamma _{p}\propto \frac{1}{\Gamma }$ for $\pi \Gamma \Delta {\gg }g_{0}$,
which is in accordance with the numerical results. Admittedly, as $\Gamma $
becomes larger, the above mentioned analysis is not a accurate, since the
frequency shift of $\omega _{p}$ will become much more complex because of
the dressing of phonons. However, from the numerical result, one can find
the analysis captures the main physics of the coupling dependent behavior.

\section{Conclusion}

\label{secvi}

In conclusion, The non-Markovian dynamics of a qubit under the decoherence
of structured environment is investigated without RWA and Markov
approximation. One point of view of the problem is the spin-boson model with
a Lorentz shaped spectral density. An alternative view is a qubit coupled to
harmonic oscillator (HO), which in turn coupled to a Ohmic environment. Two
different methods are applied and compared for this problem. One is a TRWA
method which is mainly analytical and the other one is the QUAPI method
which is a numerical scheme based on an exact methodology. The TRWA method
can be applied from the first point of view. And the QUAPI method can
applied from both points of views and called QUAPI1 and QUAPI2 respectively.
QUAPI1 only works well for large $\Gamma $. Since the memory time is too
long for the practical evaluation of QUAPI when $\Gamma $ is small. And
QUAPI2 works well for small $\Gamma $, since the non-adiabatic effect become
more important as $\Gamma $ increases, consequently smaller time-step $\Delta t$ and
more memory steps $\Delta k_{max}$ are required to obtain accurate result which also quickly
runs out the computational resources. We find that the TRWA method works
well for the whole parameter range of $\Gamma $ and show good agreement with
QUAPI1 and QUAPI2. On the other hand, we find that the decoherence of the
qubit can be reduced with increasing coupling between HO and bath, which may
be relevant to the design of quantum computer.

\section{acknowledgement}

This work was supported by the National Natural Science Foundation of China
(Grant No.10734020) and the National Basic Research Program of China (Grant
No. 2011CB922202).

\bibliographystyle{apsrev}
\bibliography{article}

\appendix

%
%

\section*{Figures Captions}

{Fig.~1:} Real and imaginary parts of the bath response function for the
Lorentzian spectral density. The memory time decreases to a finite range as $%
\Gamma $ increases.

{Fig.~2:} The population difference $P(t)$ as a function of time for the
off-resonance case ($\Delta =0.1\Omega $), the parameters are $g_{0}=0.1\Delta $ and $\pi \Gamma \Delta <g_{0}$, the QUAPI2 parameters are $N=400$, $M=2$, and $\Delta t=0.15/\Delta $.

{Fig.~3:} The population difference $P(t)$ as a function of time for the
off-resonance case ($\Delta =0.1\Omega $), the parameters are $g_{0}=0.1\Delta $ and $\pi \Gamma \Delta >g_{0}$, the QUAPI1 parameter $\Delta t=0.6/\Delta $.

{Fig.~4:} The population difference $P(t)$ as a function of time for the
on-resonance case ($\Delta =\Omega $), the parameters are $g_{0}=0.1\Delta $ and $\pi \Gamma \Delta <g_{0}$, the QUAPI2 parameters are $N=400$, $M=6$, and $\Delta t=0.3/\Delta $.

{Fig.~5:} The population difference $P(t)$ as a function of time for the
on-resonance case ($\Delta =\Omega $), the parameters are $g_{0}=0.1\Delta $ and $\pi \Gamma \Delta >g_{0}$,
the parameters are $\Delta =\Omega $, $g_{0}=0.1\Delta $ and the QUAPI1
parameter $\Delta t=0.6/\Delta $.

\clearpage
\begin{figure}[tbp]
\includegraphics[scale=0.8]{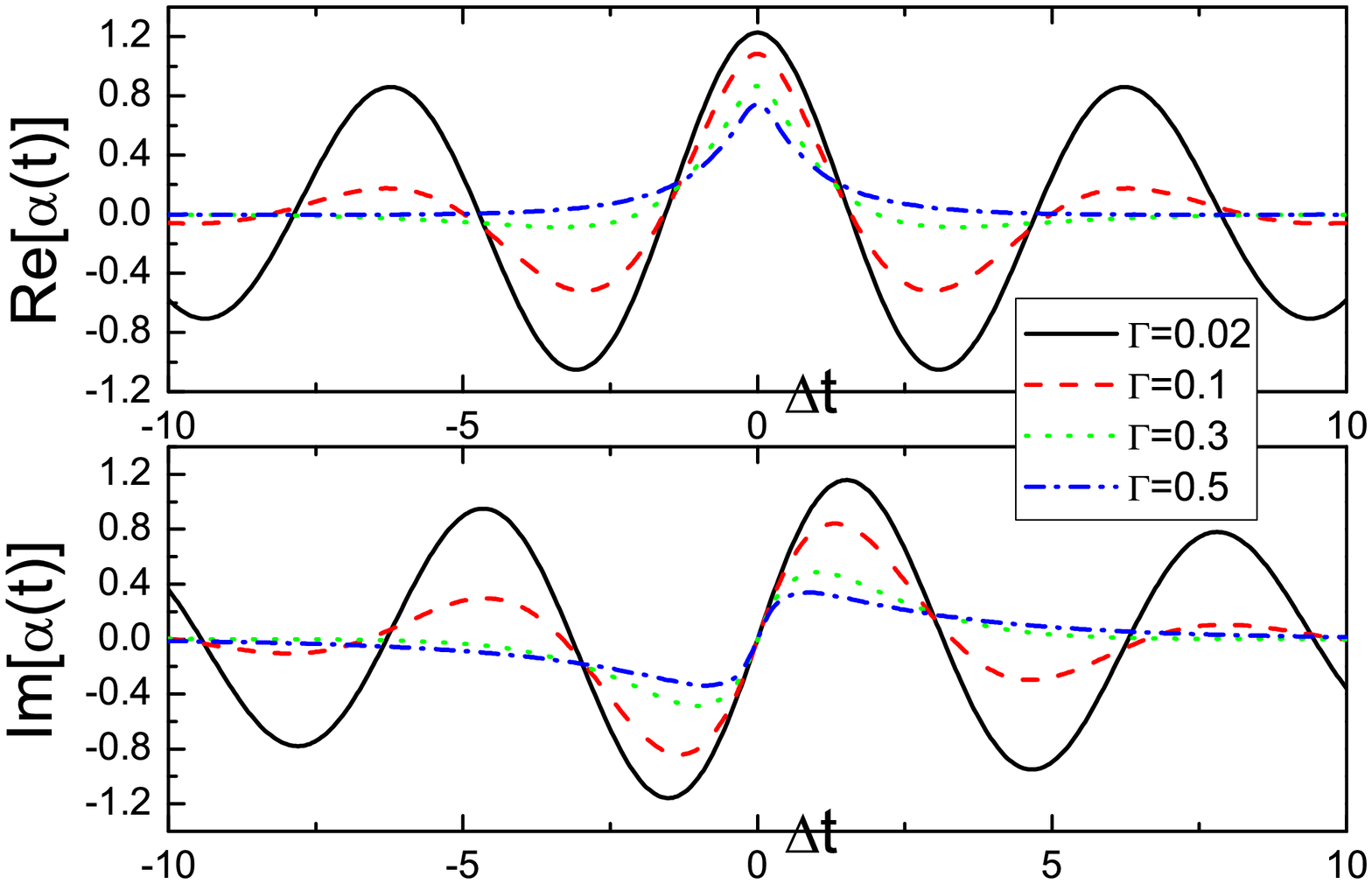}
\caption{Fig.~1}
\end{figure}
\clearpage
\begin{figure}[tbp]
\includegraphics[scale=0.8]{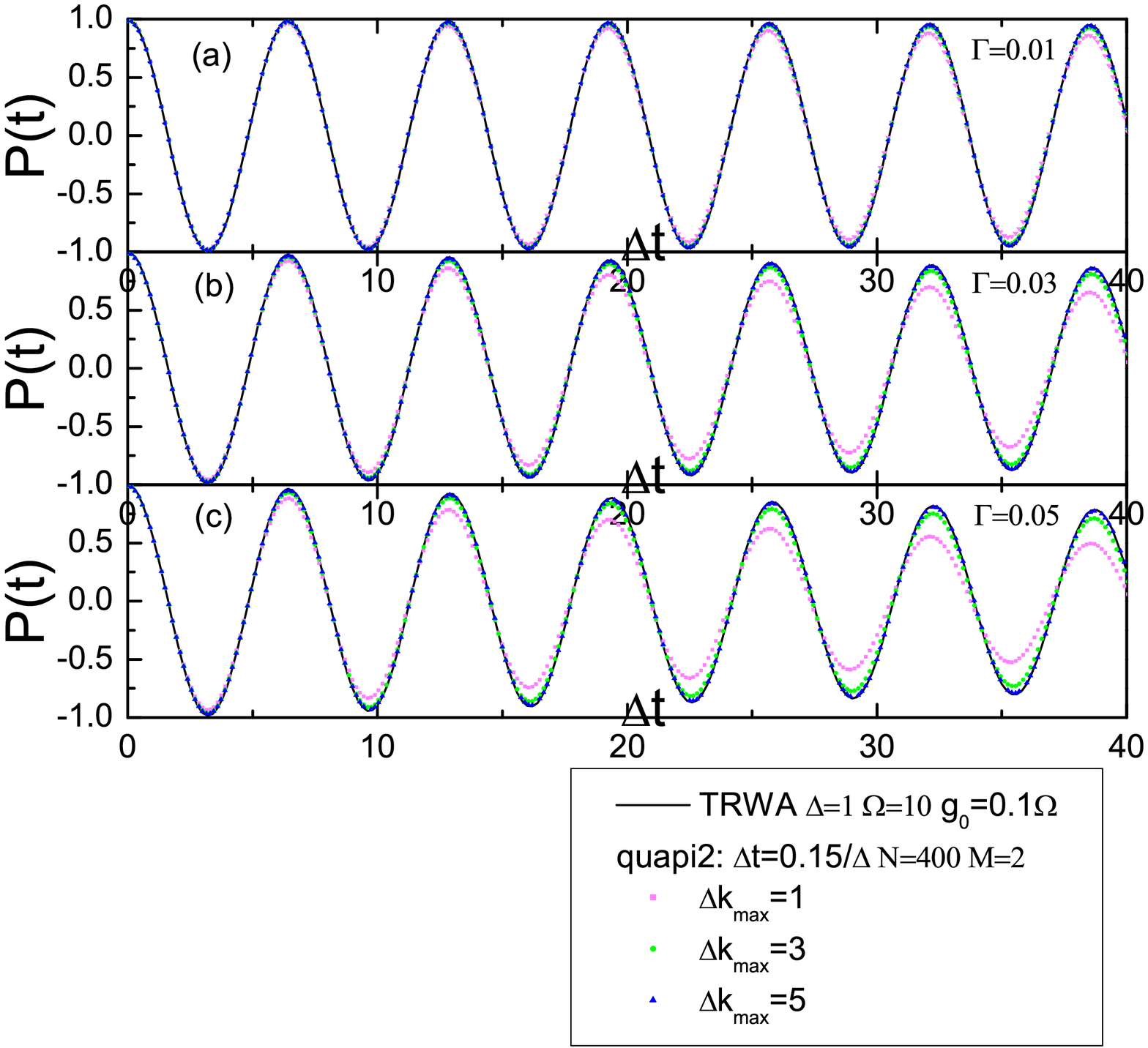}
\caption{Fig.~2}
\end{figure}
\clearpage
\begin{figure}[tbp]
\includegraphics[scale=0.8]{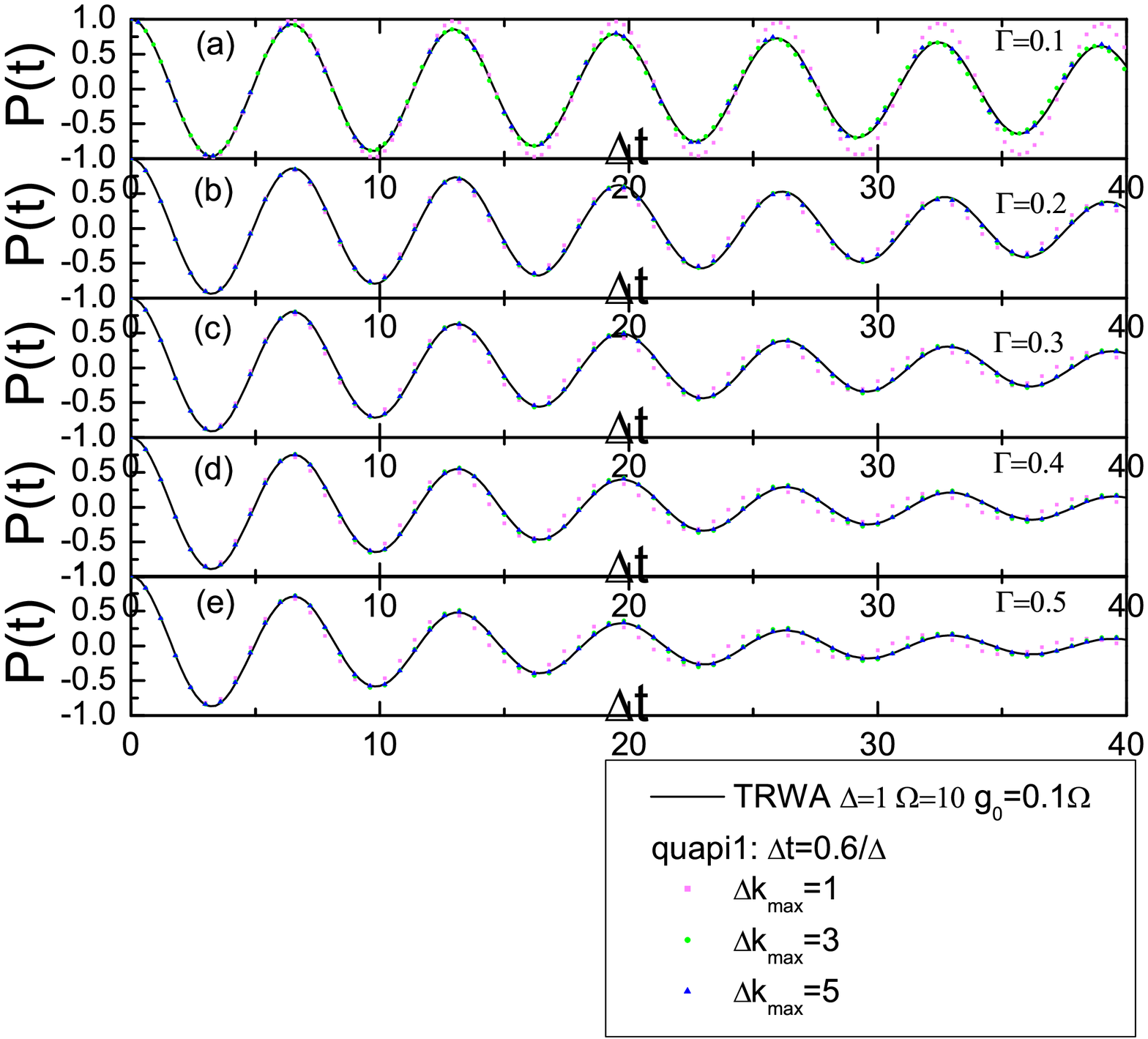}
\caption{Fig.~3}
\end{figure}
\clearpage
\begin{figure}[tbp]
\includegraphics[scale=0.8]{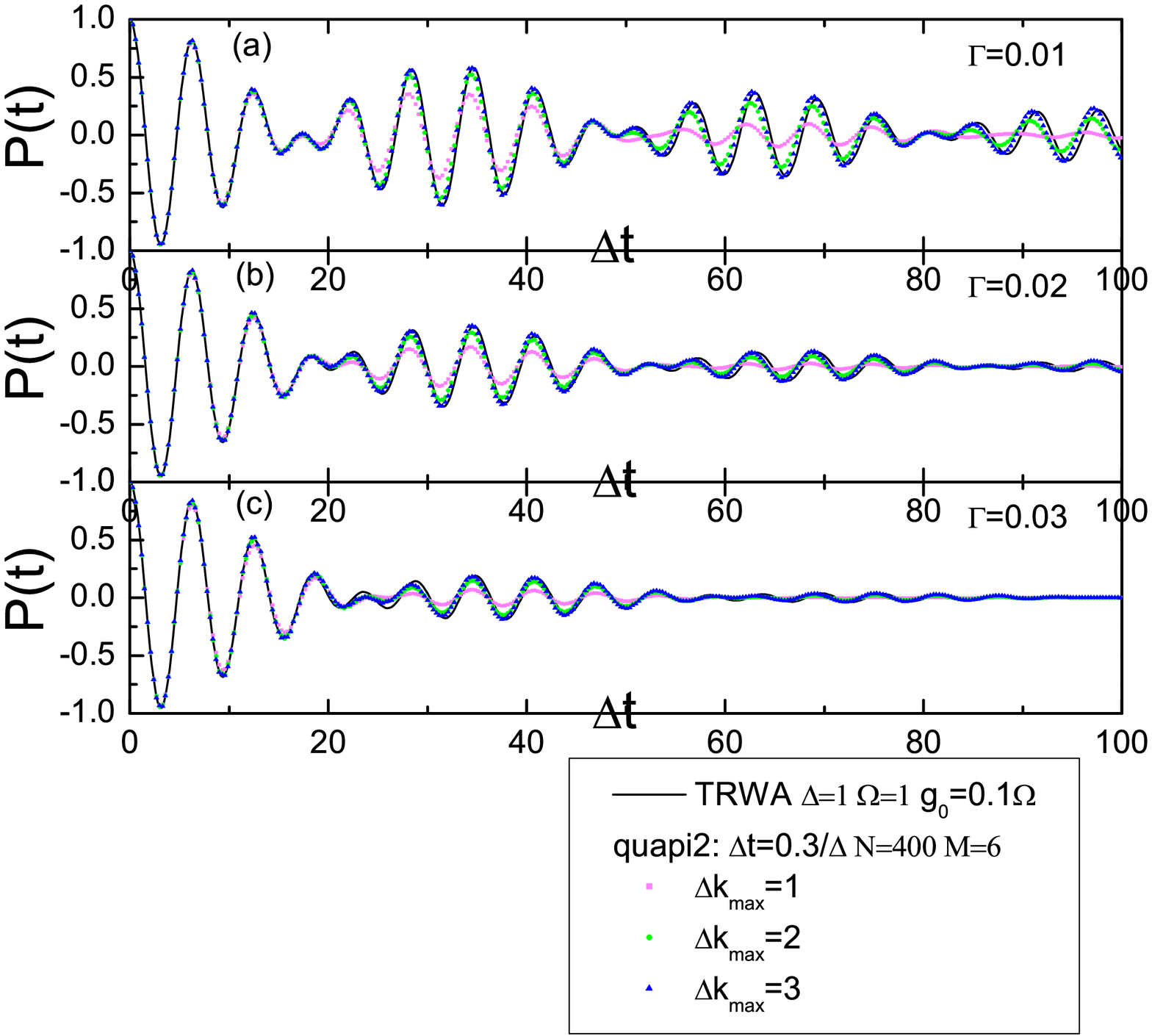}
\caption{Fig.~4}
\end{figure}
\clearpage
\begin{figure}[tbp]
\includegraphics[scale=0.8]{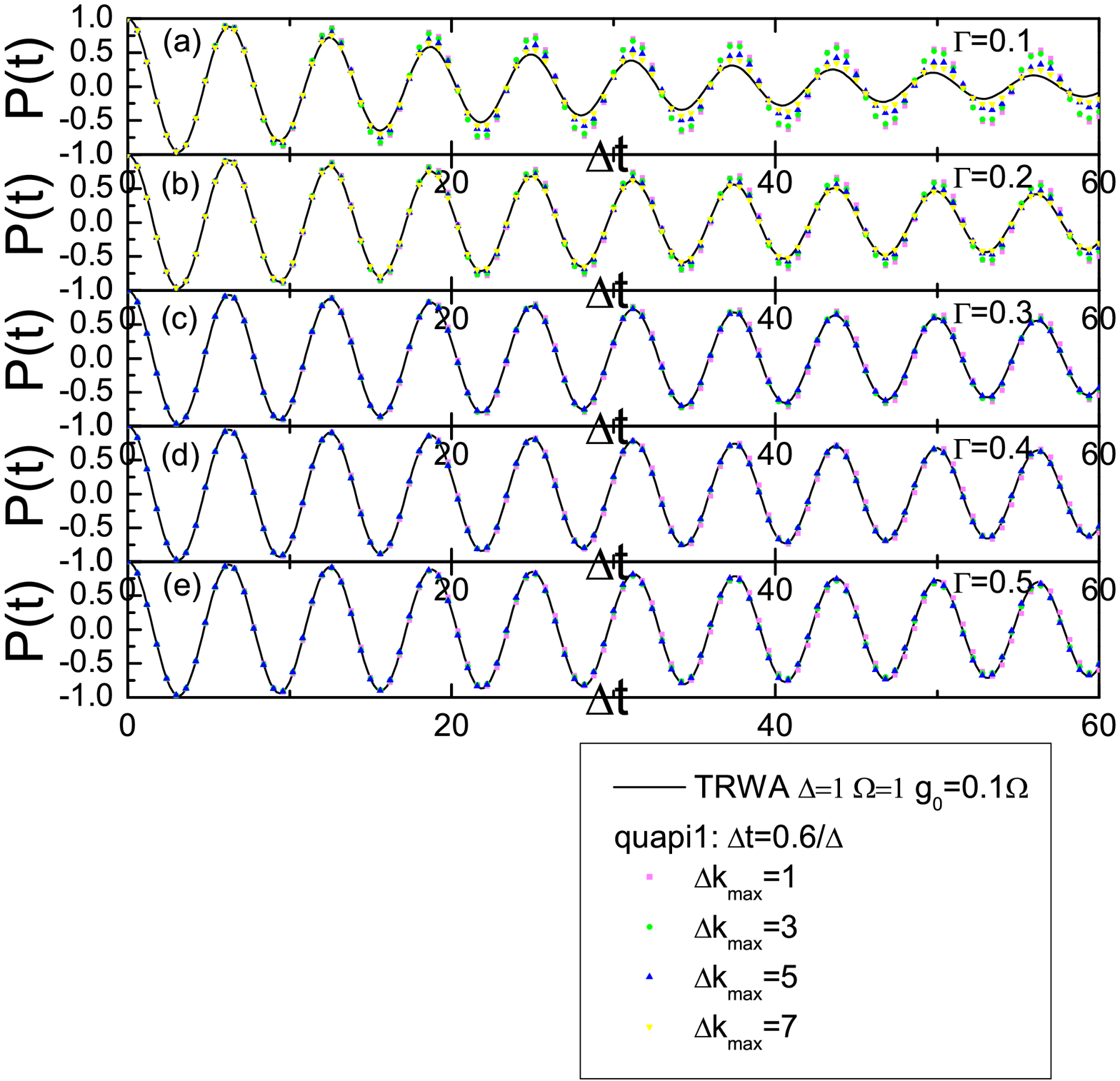}
\caption{Fig.~5}
\end{figure}

\end{document}